\documentclass{article}
\usepackage{graphicx}

\begin{document}

\title{Entanglement bounds for squeezed non-symmetric thermal state}
\author{Li-zhen JIANG , \\
Physics Department, China Institute of Metrology,\\
Hangzhou, 310034,China}
\date{}
\maketitle

\begin{abstract}
I study the three parameters bipartite quantum Gaussian state called
squeezed asymmetric thermal state, calculate Gaussian entanglement of
formation analytically and the up bound of relative entropy of entanglement,
compare them with coherent information of the state. Based on the result
obtained, one can determine the relative entropy of entanglement of \ the
state with infinitive squeezing.
\end{abstract}

\section{Introduction}

\ Quantum entanglement is one of the most important phenomenon in quantum
theory. It exhibit the nature of nonlocal correlation between quantum
systems, and plays an essential role in various fields of quantum
information processing, such as quantum computation, quantum
communication,quantum cryptography, quantum teleportation\cite
{BennettDiVincenzo},and closely related to quantum channel\cite{ChenQiu}.\
After the first experiments \cite{Furusawa} on quantum teleportation using
two-mode squeezed states \cite{Vaidman}\cite{Braunstein}, a significant
amount of work has been devoted to develop a quantum information theory of
continuous variable systems. So far, most of the theoretical work has
focused on the entanglement properties of the quantum states involved in all
these experiments, the so-called Gaussian states. The first problem arisen
is that if a given quantum Gaussian state is entangled, the problem of
qualifying entanglement has been solved in the general bipartite setting
\cite{Duan}\cite{Simon}. But the efficiency of entanglement manipulation
protocols used in practical quantum information processing critically
depends on the quality of the entanglement that one can generate. It is
therefore essential to be able to quantify the amount of entanglement in
systems with continuous variables especially Gaussian states. Several
measures have been proposed to quantify the amount of entanglement \cite
{Horodecki1}. Among which are the entanglement of formation, relative
entropy of entanglement and distillation entanglement. Entanglement of
formation is one of the most important entanglement measures. For symmetric
Guassian state, this entanglement measure can be carried out analytically
\cite{Giedke}. For general bipartite Gaussian state, another entanglement
measure called Gaussian entanglement of formation was introduced\cite{Wolf}.
Clearly it is the up bound of the entanglement of formation. Relative
entropy of entanglement\cite{Vedral1} \cite{Vedral2} on the other hand
measures the distance between the state under consideration to the closest
separable state. It has the advantage that separable states obviously
correspond to zero distance. For pure states of bipartite systems it reduces
to the von Neumann entropy of the reduced state of either subsystem. For
mixed bipartite states it is usually difficult to be\ calculated, except for
some specific states. The distance of the Gaussian state to the set of
separable Gaussian states measured by the relative entropy was considered
\cite{Scheel}. Although one has yet no proof that there does not exist a
non-Gaussian separable state which is closer to the Gaussian state under
consideration than the closest separable Gaussian state, one has good reason
to think that it is a good bound of entanglement measure. However, the
expression derived \cite{Scheel} is still not easy for a direct numeric
calculation. Recently the bound on the relative entropy of entanglement was
calculated for two modes squeezed symmetric thermal state\cite{Chen}. In
this paper I \ will consider the more practical used state of two mode
squeezed non-symmetric themal state, which is to be distributed between two
parties by means of a lossy optical fiber in asymmetric settings for an
initial two mode squeezed vacuum state\cite{DuanGuo}.I will concentrate on
its Gaussian entanglement of formation and its up bound of relative entropy
of entanglement. According to the hypothesis of hashing inequality\cite
{Horodecki2}, all this entanglement measures should be lower bounded by
coherent information. I will at last compare the results with coherent
information of the state.

\section{Squeezed non-symmetric thermal state}

Gaussian state is completely specified by its mean and its covariance
matrix, where the mean can be dropped by local unitary operation so that is
irrelevant for entanglement problems. Let us consider\ a quantum Gaussian
state $\rho $ of two system A and B acting on a Hilbert space $L^{2}\left(
\Re \right) \otimes L^{2}\left( \Re \right) $. (In quantum optical term, it
is called two modes.) It is convenient to describe Gaussian state density
operator $\rho $ by its characteristic function. (e.g. \cite{Holevo})$\chi
\left( z\right) =Tr\left( \rho V\left( z\right) \right) .$Here $z=\left(
q_{A},q_{B},p_{A},p_{B}\right) \in \Re ^{4}$ is a real vector and $V\left(
z\right) =\exp [-i\sum_{k=A,B}\left( q_{k}X_{k}+p_{k}P_{k}\right) ]$ is
\textit{Weyl operator} (displacement operator) with $X_{k}$ and $P_{k}$ are
operators of system A and B respectively, satisfying canonical commutation
relations . A characteristic function $\chi $ uniquely defines a state $\rho
_{\chi }$. Gaussian state is state whose $\chi $ is Gaussian function of $z$%
: $\chi \left( z\right) =\exp [i\eta ^{T}z-\frac{1}{4}z^{T}\gamma z]$, where
$\gamma $ is a real symmetric matrix, called correlation matrix.

Any Gaussian state of two modes can be transformed into what we called the
standard form, using local unitary operation only \cite{Duan} \cite{Simon}.
The corresponding characteristic function has displacement $\eta =0$ and the
correlation matrix has the simple form of four parameters. It was proved
that (see \cite{Holevo} and reference therein, also see \cite{Simon} and
reference therein): For arbitrary real symmetric matrix $\gamma $ there
exists linear transformation (symplectic transformation) $S_{p}:z\rightarrow
S_{p}z$ , preserving canonical commutation relations, then
\begin{equation}
\gamma \rightarrow \gamma _{0}=S_{p}\gamma S_{p}^{T}=diag\left( \gamma
_{A},\gamma _{B},\gamma _{A},\gamma _{B}\right) .
\end{equation}
This is to say that by proper symplectic transformation, Gaussian state can
be transformed into a direct product of thermal states $\rho \rightarrow
\rho _{0}=\rho _{0}^{A}\otimes \rho _{0}^{B}$. The one mode thermal state
density operators are of the forms
\begin{equation}
\rho _{0}^{\mu }=\left( 1-v_{\mu }\right) \sum_{m=0}^{\infty }v_{\mu
}^{m}\left| m\right\rangle _{\mu \mu }\left\langle m\right| ,
\end{equation}
where $v_{\mu }=\frac{N_{\mu }}{N_{\mu }+1}$, $\ N_{\mu }=\frac{1}{2}\left(
\gamma _{\mu }-1\right) $, $(\mu =A,B)$. The symplectic transformation $%
S_{p} $ induces a unitary operator $U\left( S_{p}\right) $ so that $\rho
\rightarrow U\left( S_{p}\right) \rho U^{+}\left( S_{p}\right) =\rho _{0}$
\cite{Simon}\cite{Fan}, and $\rho =U^{+}\left( S_{p}\right) \rho _{0}U\left(
S_{p}\right) $. For $\rho _{0}$ is a direct product of thermal states. It
represents the randomness side of state $\rho $. While the quantum
correlation side of $\rho $ should be caused by the two mode unitary
operators $U\left( S_{p}\right) $.

In this paper I will consider one of the special kind of Gaussian states
which can be generated from the thermal states $\rho _{0}=\rho
_{0}^{A}\otimes \rho _{0}^{B}$ with a simple form of $S_{p}$.

\begin{equation}
S_{p}=\left[
\begin{array}{ll}
\cosh r & -\sinh r \\
-\sinh r & \cosh r
\end{array}
\right] \oplus \left[
\begin{array}{ll}
\cosh r & \sinh r \\
\sinh r & \cosh r
\end{array}
\right] ,
\end{equation}
Then $\gamma =S_{p}^{-1}\gamma _{0}\left( S_{p}^{T}\right) ^{-1}$. The $%
S_{p}^{-1}$ induced unitary operator $U\left( S_{p}^{-1}\right) $ \cite{Fan}
is just the two-mode squeezed operator $S_{2}\left( r\right) $

\begin{eqnarray}
S_{2}\left( r\right) &=&\exp \left( a_{A}^{+}a_{B}^{+}\tanh r\right) \exp
\left[ -\left( a_{A}^{+}a_{A}+a_{B}^{+}a_{B}+1\right) \ln \cosh r\right]
\nonumber \\
&&\cdot \exp \left( -a_{A}a_{B}\tanh r\right) .
\end{eqnarray}
This kind of Gaussian states will be called squeezed non-symmetric thermal
states. And in the rest of this paper, I will use $\rho $ to specify the
states.

\begin{equation}
\rho =S_{2}\left( r\right) \rho _{0}\left( v_{A},v_{B}\right)
S_{2}^{+}\left( r\right) ,
\end{equation}
Denote $\lambda =\tanh r$, then the inseparability criterion reads

\begin{equation}
\lambda >\sqrt{v_{A}v_{B}}.
\end{equation}

For the convenience of further application, one can express $\rho $ in
coherent state representation as

\begin{eqnarray}
\left\langle \alpha _{A},\alpha _{B}\right| \rho \left| \beta _{A},\beta
_{B}\right\rangle &=&C_{0}\exp [-\frac{1}{2}\left( \left| \alpha _{A}\right|
^{2}+\left| \alpha _{B}\right| ^{2}+\left| \beta _{A}\right| ^{2}+\left|
\beta _{B}\right| ^{2}\right) ]  \nonumber \\
&&\cdot \exp [\tau \left( \alpha _{A}^{\ast }\alpha _{B}^{\ast }+\beta
_{A}\beta _{B}\right) +\omega _{A}\alpha _{A}^{\ast }\beta _{A}+\omega
_{B}\alpha _{B}^{\ast }\beta _{B}]
\end{eqnarray}
where $C_{0}=\frac{\left( 1-v_{A}\right) (1-v_{B})\left( 1-\lambda
^{2}\right) }{1-v_{A}v_{B}\lambda ^{2}}$, $\tau =\frac{\lambda \left(
1-v_{A}v_{B}\right) }{1-v_{A}v_{B}\lambda ^{2}}$ , $\omega _{A}=\frac{%
v_{A}\left( 1-\lambda ^{2}\right) }{1-v_{A}v_{B}\lambda ^{2}}$, $\ \omega
_{B}=\frac{v_{B}\left( 1-\lambda ^{2}\right) }{1-v_{A}v_{B}\lambda ^{2}}$.
With coherent state representation the reduced state can be easily obtained
by integration. The reduced density matrix $\rho ^{A}=Tr_{B}\rho $ and $\rho
^{B}=Tr_{A}\rho $ turn out to be one mode thermal states $\rho ^{\mu
}=\left( 1-v_{\mu }^{rd}\right) \sum_{m=0}^{\infty }(v_{\mu
}^{rd})^{m}\left| m\right\rangle _{\mu \mu }\left\langle m\right| $, $(\mu
=A,B)$ with parameters $v_{A}^{rd}$ and $v_{B}^{rd}$ respectively,
\begin{equation}
v_{A}^{rd}=\frac{v_{A}\left( 1-v_{B}\right) +\lambda ^{2}\left(
1-v_{A}\right) }{1-v_{B}+\lambda ^{2}\left( 1-v_{A}\right) },\quad
v_{B}^{rd}=\frac{v_{B}\left( 1-v_{A}\right) +\lambda ^{2}\left(
1-v_{B}\right) }{1-v_{A}+\lambda ^{2}\left( 1-v_{B}\right) },\quad
\end{equation}

\section{Gaussian entanglement of formation}

Entanglement of formation is defined as an infimum
\begin{equation}
E_{F}\left( \rho \right) =\inf \left\{ \sum_{k}p_{k}E\left( \Psi _{k}\right)
\left| \rho =\sum_{k}p_{k}\left| \Psi _{k}\right\rangle \left\langle \Psi
_{k}\right| \right. \right\}  \label{wave1}
\end{equation}
over all (possibly continuous) convex decompositions of the state into pure
states with respective entanglement being von Neumann entropy of the reduced
state. By its definition calculating $E_{F}$ is a highly non-trivial
optimization problem, which becomes numerically intractable very rapidly if
we increase the dimensions of the Hilbert spaces. Remarkably, there exist
analytical expressions for two-qubit systems as well as for highly symmetric
states. Recently, $E_{F}$ was calculated for symmetric Gaussian states of
two modes\cite{Giedke}. But for general two mode Gaussian states, it is
still not easy if not impossible to carry out the $E_{F}$. For these reasons
the Gaussian entanglement of formation (GEoF) $E_{G}$ is introduced \cite
{Wolf} to quantify the entanglement of bipartite Gaussian states by taking
the infimum in \ref{wave1} only over decompositions into pure Gaussian
states.

For any two-mode Gaussian state with correlation matrix $\gamma $ = $\gamma
_{q}\oplus \gamma _{p}$, It is proved \cite{Wolf} that the Gaussian
entanglement of formation is given by the entanglement of the least
entangled pure state with $\gamma _{pure}$=$X\oplus X^{-1}$ which is such
that \

\begin{equation}
\det \left( X-\gamma _{q}\right) =\det \left( X^{-1}-\gamma _{p}\right) =0.
\label{wave2}
\end{equation}

As a part of the correlation matrix of the entangled pure state, the
symmetric $2\times 2$ matrix $X$ can always be written in the form of

\begin{equation}
X=\left[
\begin{array}{ll}
yx\cosh r_{g} & y\sinh r_{g} \\
y\sinh r_{g} & yx^{-1}\cosh r_{g}
\end{array}
\right] ,
\end{equation}
then the pure bipartite state with correlation matrix $\gamma _{pure}$ can
be constructed by successively applying two local unitary operations to the
two mode squeezed vacuum state, where the two local unitary operations are $%
S_{p1}=diag\{\sqrt{y},\sqrt{y},\sqrt{y^{-1}},\sqrt{y^{-1}}\}$ and $%
S_{p2}=diag\{\sqrt{x},\sqrt{x^{-1}},\sqrt{x^{-1}},\sqrt{x}\}$ respectively
in simplectic form.

Denote $n=\left( N_{1}+N_{2}+1\right) \cosh 2r+N_{1}-N_{2}$, $m=\left(
N_{1}+N_{2}+1\right) \cosh 2r-N_{1}+N_{2}$, $k=\left( N_{1}+N_{2}+1\right)
\sinh 2r$. One has
\[
\gamma _{q}=\left[
\begin{array}{ll}
n & k \\
k & m
\end{array}
\right] \quad \gamma _{p}=\left[
\begin{array}{ll}
n & -k \\
-k & m
\end{array}
\right] \quad
\]

Eq. (\ref{wave2}) then can be written as

\begin{eqnarray}
\left( n/x+mx\right) \cosh r_{g}-2k\sinh r_{g}-[(nm-k^{2})/y+y] &=&0,
\label{wave3} \\
\left( nx+m/x\right) \cosh r_{g}-2k\sinh r_{g}-[(nm-k^{2})y+1/y] &=&0.
\label{wave4}
\end{eqnarray}

So that one has $\left( n+m\right) \left( 1/x+x\right) \cosh r_{g}-4k\sinh
r_{g}-(nm-k^{2}+1)(1/y+y)=0.$ Clearly $r_{g}$ is a monotonical increase
function of $(1/y+y)$, and when $y=1$, the minimal value of $r_{g}$ will be
achieved. One can subsequently obtains $x=1$ by substracting Eq.(\ref{wave3}%
) from Eq.(\ref{wave4}). The Gaussian entanglement of formation $E_{G}$%
\bigskip\ will be
\begin{equation}
E_{G}=g(\sinh ^{2}r_{g}).
\end{equation}
where $g(x)=(x+1)\log (x+1)-x\log x$ is the bosonic entropy function, \ and

\begin{equation}
r_{g}=r-\frac{1}{2}\ln \frac{1+\sqrt{v_{A}v_{B}}}{1-\sqrt{v_{A}v_{B}}}.
\end{equation}
The second term at the right hand side can be written of as $r_{0},$ with $%
\sqrt{v_{A}v_{B}}=\tanh r_{0}$, representing the noise side of the bipartite
state. While $r$ represents quantum correlation side of the state. The
difference of the two gives the GEoF squeezing parameter. The compaison of
the GEoFs for different ratio of average photon numbers $N_{B}/N_{A}$ \ is
displayed \bigskip in \ Fig. 1.

\section{\protect\bigskip Up bound for relative entropy of entanglement}

The relative entropy of entanglement for bipartite quantum state $\sigma $
is defined by\cite{Vedral1}:

\begin{equation}
E_{r}\left( \sigma \right) \equiv \min_{\widetilde{\sigma }\in D}S\left(
\sigma \left\| \widetilde{\sigma }\right. \right)
\end{equation}
where $D$ is the set of all disentangled states, and $S\left( \sigma \left\|
\widetilde{\sigma }\right. \right) \equiv Tr\left\{ \sigma \left( \log
\sigma -\log \widetilde{\sigma }\right) \right\} $ is the relative entropy
of $\sigma $ with respect to $\widetilde{\sigma }$. Consider the relative
entropy of entanglement of squeezed thermal state $\rho $, if one chooses a
subset of $D$ which contain all Gaussian separable state, or more
specifically\ all separable squeezed thermal state to substitute the
district $D$ to carry out the minimum of relative entropy, clearly such
minimums are local minimums. They can be utilized as up bound $E_{ur}\left(
\rho \right) $ of global minimum, the relative entropy of entanglement $%
E_{r}\left( \rho \right) $. Denote the district of all separable squeezed
thermal states as $D_{ST}$, then $D_{ST}\subset D$, and $E_{ur}\left( \rho
\right) \leq E_{r}\left( \rho \right) $, generally speaking the identity can
not be achieved. Even for relative entropy of pure Gaussian state, the
minimum is achieved by non-gaussian separable state \cite{Vedral2}. Clearly
the above way of obtaining up bound by shrinking the district of
minimization can be applied to other Gaussian state or any other state. In
order to obtain the up bound of the relative entropy of entanglement of
squeezed thermal state $\rho $, Let us first consider the relative entropy
of $\rho $ with respect to seperable squeezed thermal state $\widetilde{\rho
}(\widetilde{v}_{A},\widetilde{v}_{B})=S_{2}\left( \widetilde{r}\right) \rho
_{0}\left( \widetilde{v}_{A},\widetilde{v}_{B}\right) S_{2}^{+}\left(
\widetilde{r}\right) $, with $\widetilde{\lambda }=\tanh \widetilde{r}\leq
\sqrt{\widetilde{v}_{A}\widetilde{v}_{B}}$. The von Neumann entropy of state
$\rho $ is \cite{Holevo}
\begin{equation}
S\left( \rho \right) =-Tr\rho \log \rho =g\left( N_{A}\right) +g(N_{B})
\end{equation}
with $N_{\mu }=\frac{v_{\mu }}{1-v_{\mu }},$ \ $(\mu =A,B)$. And one gets $%
\log \widetilde{\rho }=S_{2}\left( \widetilde{r}\right) (\log \rho
_{0}\left( \widetilde{v}_{A},\widetilde{v}_{B}\right) )S_{2}^{+}\left(
\widetilde{r}\right) $ by the unitary of $S_{2}\left( \widetilde{r}\right) $%
. Put $\rho _{0}(\widetilde{v})$\ in explicit operator form,
\begin{equation}
\rho _{0}\left( \widetilde{v}\right) =\left( 1-\widetilde{v}_{A}\right) (1-%
\widetilde{v}_{B})\widetilde{v}_{A}^{a_{A}^{+}a_{A}\ }\widetilde{v}%
_{B}^{a_{B}^{+}a_{B}}.
\end{equation}
Then the second part of relative entropy will be

\begin{eqnarray}
Tr\rho \log \widetilde{\rho } &=&\log \left( 1-\widetilde{v}_{A}\right)
+\log (1-\widetilde{v}_{B})+Tr\{S_{2}\left( r\right) \rho _{0}\left(
v_{A},v_{B}\right) S_{2}^{+}\left( r\right)  \nonumber \\
&&S_{2}\left( \widetilde{r}\right) \left( \log \widetilde{v}%
_{A}a_{A}^{+}a_{A}+\log \widetilde{v}_{B}a_{B}^{+}a_{B}\right)
S_{2}^{+}\left( \widetilde{r}\right) \}
\end{eqnarray}
By utilizing $S_{2}\left( r\right) a_{A}S_{2}^{+}\left( r\right) =a_{A}\cosh
r-a_{B}^{+}\sinh r$ and with the property that operator can be cycled under
the trace, after some algebra one obtains

\begin{eqnarray}  \label{wave}
Tr\rho \log \widetilde{\rho } &=&\log \left( 1-\widetilde{v}_{A}\right)
+\log (1-\widetilde{v}_{B})  \nonumber \\
&&+[\frac{v_{A}}{1-v_{A}}\cosh ^{2}\left( r-\widetilde{r}\right) +\frac{1}{%
1-v_{B}}\sinh ^{2}\left( r-\widetilde{r}\right) ]\log \widetilde{v}_{A}
\nonumber \\
&&+[\frac{v_{B}}{1-v_{B}}\cosh ^{2}\left( r-\widetilde{r}\right) +\frac{1}{%
1-v_{A}}\sinh ^{2}\left( r-\widetilde{r}\right) ]\log \widetilde{v}_{B}
\end{eqnarray}
where $Tr[(a_{A}^{+}a_{B}^{+}+a_{A}a_{B})\rho _{0}(v_{A},v_{B})]=0$ is
applied. Let us first find out the maximum point of Eq. (\ref{wave}) by
partial differentiation with respect to $\widetilde{r}$ and $\widetilde{v}%
_{A},\widetilde{v}_{B}$ regardless the fact that $\widetilde{\lambda }=\tanh
\widetilde{r}\leq \sqrt{\widetilde{v}_{A}\widetilde{v}_{B}}$, it will be at
the point $\left( \widetilde{r},\widetilde{v}_{A},\widetilde{v}_{B}\right)
=\left( r,v_{A},v_{B}\right) $. Meanwhile it is noticed that there is no
other maximum point. Then let us add the condition of $\widetilde{\lambda }%
\leq \sqrt{\widetilde{v}_{A}\widetilde{v}_{B}}$, the maximum should be
achieved at the edge of $D_{ST}$, that is $\widetilde{\lambda }=\sqrt{%
\widetilde{v}_{A}\widetilde{v}_{B}}$. After one of the parameter is
determined, the remain problem is to seek out the maximum with respect to $%
\widetilde{v}_{A},\widetilde{v}_{B}$. If the maximum is achieved at $(%
\widetilde{v}_{A},\widetilde{v}_{B})=(\widetilde{v}_{A}^{\ast },\widetilde{v}%
_{B}^{\ast })$, and denote $\sqrt{\widetilde{v}_{A}^{\ast }\widetilde{v}%
_{B}^{\ast}}=\tanh \widetilde{r}^{\ast }$, The up bound of relative entropy
of entanglement for squeezed thermal state will be

\begin{eqnarray}
E_{ur}\left( \rho \right) &=&-g\left( N_{A}\right) -g(N_{B})-\log \left( 1-%
\widetilde{v}_{A}^{\ast }\right) +\log (1-\widetilde{v}_{B}^{\ast })
\nonumber \\
&&+[\frac{v_{A}}{1-v_{A}}\cosh ^{2}\left( r-\widetilde{r}^{\ast }\right) +%
\frac{1}{1-v_{B}}\sinh ^{2}\left( r-\widetilde{r}^{\ast }\right) ]\log
\widetilde{v}_{A}^{\ast }  \nonumber \\
&&+[\frac{v_{B}}{1-v_{B}}\cosh ^{2}\left( r-\widetilde{r}^{\ast }\right) +%
\frac{1}{1-v_{A}}\sinh ^{2}\left( r-\widetilde{r}^{\ast }\right) ]\log
\widetilde{v}_{B}^{\ast }
\end{eqnarray}

\section{Hashing inequality and comparison of the \newline
bounds with coherent information}

One of the most concerned problem in quantum information is the quantum
capacities, it is shown that the rate of quantum information transformation
is bounded by the maximal attainable rate of coherent information. If there
exists hashing inequality, that is, for any bipartite state the one-way
distillable entanglement is no less than coherent information, then one
obtains Shannon-like formulas for the capacities \cite{Horodecki2}. The
coherent information of bipartite state $\sigma $ with reductions $\sigma
_{A}$ and $\sigma _{B}$ is defined as \cite{Schumacher}

\begin{equation}
I^{\mu }\left( \sigma \right) =S\left( \sigma _{\mu }\right) -S\left( \sigma
\right) ,
\end{equation}
for $S\left( \sigma _{\mu }\right) -S\left( \sigma \right) \geq 0$ and $%
I^{\mu }\left( \sigma \right) =0$ otherwise. The hypothetical hashing
inequality mentioned above is

\begin{equation}
D_{\rightarrow }\left( \sigma \right) \geq I^{\mu }\left( \sigma \right)
\end{equation}
where $D_{\rightarrow }\left( \sigma \right) $ is forward classical
communication aided distillable entanglement of the state. We know that
relative entropy of entanglement $E_{r}\left( \sigma \right) $ is no less
than two way distillation of entanglement $D_{\leftrightarrow }\left( \sigma
\right) $\cite{Vedral3} , the later is no less than one way distillation of
entanglement $D_{\rightarrow }\left( \sigma \right) $. Combining with
hashing inequality\ one has

\begin{equation}
E_{ur}\left( \sigma \right) \geq E_{r}\left( \sigma \right) \geq
D_{\leftrightarrow }\left( \sigma \right) \geq D_{\rightarrow }\left( \sigma
\right) \geq I^{B}\left( \sigma \right) .
\end{equation}
One the other hand, it is proved that relative entropy is lower bounded by
coherent information\cite{Plenio}, so $E_{r}\left( \sigma \right) $ $\geq
I^{B}\left( \sigma \right) $ is always true irrespective of the hypothesis
of hashing inequality.

Consider the squeezed non-symmetric thermal state $\rho $, the reduced
states $\rho ^{A}=Tr_{B}\rho $ and $\rho ^{B}=Tr_{A}\rho $ turn out to be\
not the same. The entropies of its reduced states are $g(N_{A}^{rd})$ and $%
g(N_{B}^{rd})$ respectively, where $\ N_{\mu }^{rd}=v_{\mu }^{rd}/(1-v_{\mu
}^{rd})$ is the average particle number of the reduced state. This leads to
two different coherent information $I^{A}(\rho )$ and $I^{B}\left( \rho
\right) $. They are

\begin{equation}
I^{\mu }\left( \rho \right) =\max \{g(N_{\mu }^{rd})-g(N_{A})-g(N_{B}),0\}
\end{equation}

The numerical results are shown in the Fig. 2.

\section{Conclusions}

A subset of quantum Gaussian mixed state is given and scrutinizingly
investigated. The so called squeezed non-symmetric thermal state has several
merits: It can be described by three parameters, one for quantum
correlation, the other two for randomness; The separability of the state
readily follows from comparison of the first parameter and the geometric
average of the last two parameters, if the randomness is stronger, the state
is separable, otherwise it is entangled; All the three parameters have range
from zero to one. It is the simplest Gaussian mixed state other than the two
parameters squeezed thermal state. The $\lambda ,v_{A},v_{B}$ description of
state presented in this paper enable the calculation more comparable. \ With
the local unitary operation method, I obtain the GEoF analytically and write
it in a simple style with clearly physical meaning.  By the nemeric results
I find that: As the squeezing parameter tends towards infinitive, the up
bound of relative entropy of entanglement tend to coincide with the coherent
information, so that one can determine the relative entropy of entanglement,
and if the hashing inequality is right, the distillation entanglement can be
determined at infinitive squeezing.

\begin{figure}[tbp]
\begin{center}
\includegraphics[
trim=0.000000in 0.000000in -0.138042in 0.000000in,
height=2.0081in, width=2.5097in ]{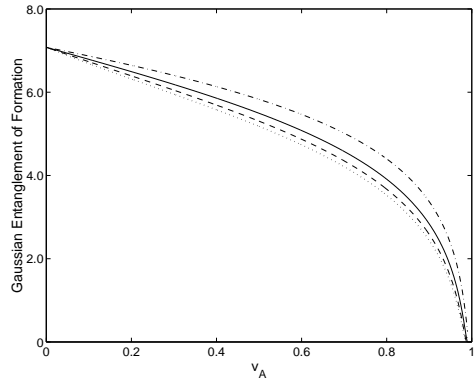}
\end{center}
\caption{Comparison of GEoFs. With $\protect\lambda$ =0.99,
dash-dot,solid,dash,and dot line for $N_{B}/N_{A}$ =0.5,1.0,1.5,2.0
respectively.}
\end{figure}

\begin{figure}[tbp]
\begin{center}
\includegraphics[
trim=0.000000in 0.000000in -0.138042in 0.000000in,
height=2.0081in, width=2.5097in ]{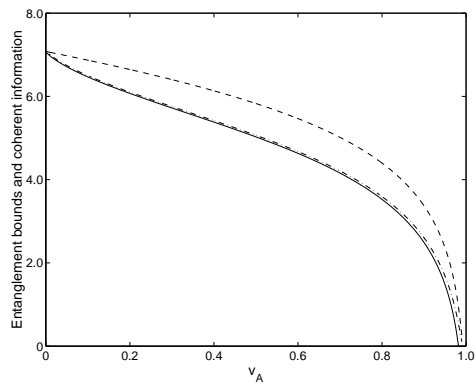}
\end{center}
\caption{Comparison of the GEoF, up bound relative entropy of entanglement
of with coherent information. With $\protect\lambda =0.99,$ and $%
N_{B}/N_{A}=0.5$, dot-dash for $E_{ur}$, solid for $I^{A}$, and dash line
for GEoF. }
\end{figure}

\end{document}